\documentclass[aps,prl,twocolumn,superscriptaddress, amsmath, amssymb]{revtex4-1}

\usepackage[latin1]{inputenc}
\usepackage{graphicx}
\usepackage{dcolumn}
\usepackage{bm}
\usepackage{color}
\usepackage{tabularx}
\newcolumntype{Y}{>{\centering}X}
\usepackage[cdot,textstyle,amssymb]{SIunits}

\usepackage{sistyle}
\SIunitspace{\cdot}
\SIunitdot{\cdot}
\SIdecimalsign{.}

\usepackage[pdfborder={0 0 0}, colorlinks=true,citecolor=blue,linkcolor=black,urlcolor=blue]{hyperref}
\begin{document}


\title{Large Thermal Hall Effect in $\alpha$-RuCl$_3$: Evidence for Heat Transport by Kitaev-Heisenberg Paramagnons}





\author{Richard Hentrich}
 \email{r.hentrich@ifw-dresden.de}
\affiliation{Leibniz Institute for Solid State and Materials Research, 01069 Dresden, Germany}
%
%
%

\author{Maria Roslova}
\affiliation{Faculty of Chemistry and Food Chemistry, TU Dresden, 01062 Dresden, Germany}
\author{Anna Isaeva}
\affiliation{Faculty of Chemistry and Food Chemistry, TU Dresden, 01062 Dresden, Germany}
\author{Thomas Doert}
\affiliation{Faculty of Chemistry and Food Chemistry, TU Dresden, 01062 Dresden, Germany}

\author{Wolfram Brenig}
\affiliation{Institute for Theoretical Physics, TU Braunschweig, 38106 Braunschweig, Germany}

%
%
%
%
%
%
%
%

\author{Bernd B\"uchner }
\affiliation{Leibniz Institute for Solid State and Materials Research, 01069 Dresden, Germany}
\affiliation{Institute of Solid State Physics, TU Dresden, 01069 Dresden, Germany}
\affiliation{Center for Transport and Devices, TU Dresden, 01069 Dresden, Germany}

\author{Christian Hess}\email{c.hess@ifw-dresden.de}
\affiliation{Leibniz Institute for Solid State and Materials Research, 01069 Dresden, Germany}
\affiliation{Center for Transport and Devices, TU Dresden, 01069 Dresden, Germany}


\date{\today}


\begin{abstract}
The honeycomb Kitaev model in a
magnetic field is a source of a topological quantum spin liquid with Majorana
fermions and gauge flux excitations as fractional quasiparticles.  We present
experimental results for the thermal Hall effect of the material $\alpha$-RuCl$_{3}$
which recently emerged as a prime candidate for realizing such physics. At
temperatures above long-range magnetic ordering $T\gtrsim T_N\approx8$~K, we observe
with an applied magnetic field $B$ perpendicular to the honeycomb layers a sizeable
positive transversal heat conductivity $\kappa_{xy}$ which increases linearly with
$B$. Upon raising the temperature, $\kappa_{xy}(T)$ increases strongly, exhibits a broad maximum at around 30~K, and
eventually becomes negligible at $T\gtrsim 125$~K. Remarkably, the longitudinal heat
conductivity $\kappa_{xx}(T)$ exhibits a sizeable positive thermal magnetoresistance
effect. Thus, our findings provide clear-cut evidence for longitudinal and
transverse magnetic heat transport and underpin the unconventional nature of the quasiparticles in the paramagnetic phase of $\alpha$-RuCl$_{3}$.

\end{abstract}



\maketitle

About ten years ago, Kitaev proposed a new type of quantum spin model whose ground state has been exactly shown to be a realization of a gapless
quantum spin-liquid (QSL)  \cite{Kitaev2006}. This is a peculiar state of matter where
magnetic long-range order is suppressed due to substantial quantum fluctuations even
at zero temperature  \cite{Lee2008,Balents2010,Savary2017}. In the presence of
certain time-reversal-symmetry breaking perturbations, e.g. external magnetic fields, this
spin liquid opens a gap, leading to a topologically non-trivial spin liquid
(TQSL).  The striking feature of Kitaev's TQSL is, that in addition to its
unconventional bulk excitations, which are due to fractionalization of spins into
localized $Z_2$ gauge fluxes and itinerant Majorana fermions  \cite{Kitaev2006,
Baskaran2007, Knolle2014}, and which are present already in the gapless state, a {\it
chiral} Majorana edge mode arises in the field induced gap and the Z$_2$ vortices
acquire non-Abelian anyonic statistics  \cite{Kitaev2006}.

When investigating possible candidate materials for realizing such exotic physics, heat conductivity experiments are considered one of the few probes to study the TQSL quasiparticle fingerprints because information on the quasiparticles' specific heat, their velocity, and their scattering is provided \cite{Qi2009}.
The magnetically frustrated honeycomb compound $\alpha$-RuCl$_3$ has been intensively studied recently as it has been suggested to host a proximate Kitaev TQSL \cite{Banerjee2016,Banerjee2017}. 
A profound understanding of how putative fractional magnetic excitations contribute to the heat transport in $\alpha$-RuCl$_3$ is thus highly desirable. 

\begin{figure}[t]
\centering
\includegraphics[width=\columnwidth]{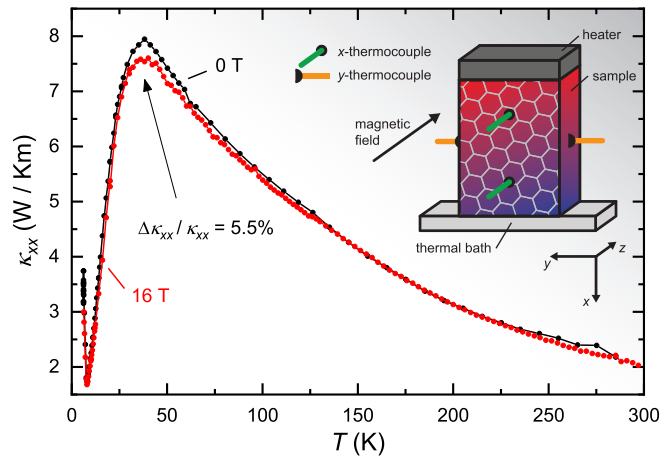}
\caption{\label{kapx}
Temperature dependence of the longitudinal heat conductivity $\kappa_{xx}$ of $\alpha$-RuCl$_3$ parallel to the honeycomb layers at zero field and at a magnetic field $B_z= 16$~T perpendicular to the layers. The inset shows a sketch of the experimental setup.}
\end{figure}

Several heat conductivity studies on $\alpha$-RuCl$_3$ yield an inconsistent picture on possible heat transport by emergent quasiparticles of the spin system. 
Heat transport by itinerant spin excitation has been inferred from an anomaly in the in-plane longitudinal heat conductivity $\kappa_{xx}$ at around 100~K \cite{Hirobe2017} and from a magnetic field-induced low-temperature enhancement of $\kappa_{xx}$ for fields $B\gtrsim8$~T parallel to the material's honeycomb planes \cite{Leahy2017}. These interpretations however, have recently been ruled out 
by results for the out-of-plane heat conductivity $\kappa_{zz}$ \footnote{We refer to an orthogonal coordinate system with the $x$ and $y$ directions within and $z$ perpendicular to the planes of $\alpha$-RuCl$_3$, see inset of Fig.~\ref{kapx}.} where both types of anomalies are present as well \cite{Hentrich2018}. The emergent picture for rationalizing the heat transport in this material is thus that its most salient features can be explained by phononic heat transport \cite{Hentrich2018,Yu2018}. However, it remains to be settled whether a finite magnetic contribution to the heat transport, possibly obscured by the phononic transport, is present in the material.

A natural way to prove the presence of low-dimensional magnetic heat transport is the systematic analysis of the anisotropy of the heat conductivity tensor. This approach has successfully been applied e.g. for low-dimensional quantum magnets realized in cuprates \cite{Sologubenko00,Hess01,Hess03,Hess04a,Hess2007,Hess2007b,Hlubek2010}. For $\alpha$-RuCl$_3$, however, the anisotropy of the heat conductivity tensor is relatively 
weak \cite{Hentrich2018}, which renders it difficult to draw clear-cut conclusions with respect to the presence of magnetic heat transport.
An alternative way to unambiguously prove such transport is the investigation of the thermal Hall effect, i.e. the transversal heat conductivity $\kappa_{xy}$ in an external magnetic field. Originally recognized to be relevant for conduction electrons in metals (the so-called Righi-Leduc effect \cite{Ziman}), it has recently been noticed theoretically that the thermal Hall effect represents a viable tool for detecting the transport by magnetic quasiparticles of chiral spin systems and spin liquids \cite{Fujimoto2009,Katsura2010}. Indeed, clear-cut evidence for a finite magnon thermal Hall effect has been reported for several classical chiral magnet and quantum spin ice systems \cite{Onose2010,Ideue2012,Hirschberger2015}.

Here, we report a sizeable thermal Hall effect for $\alpha$-RuCl$_3$.
In this compound, strong spin-orbit coupling and an edge-sharing configuration of RuCl$_{6}$ octahedra yield a honeycomb lattice of  $j_\mathrm{eff}=1/2$ states with dominant Kitaev interaction \cite{Plumb2014,Koitzsch2016,Banerjee2016,Yadav2016}. Long-range magnetic order at $T_N\approx7$~K occurs in as-grown samples of $\alpha$-RuCl$_3$ without stacking faults, whereas stacking disorder causes a spread of the ordering temperature to up to about 14~K \cite{Sears2015,Kubota2015,Cao2016}. While a moderate in-plane magnetic field of $\sim8$~T is sufficient to completely suppress the long-range magnetic order, the latter persists in standard laboratory fields perpendicular to the honeycomb layers \cite{Johnson2015,Majumder2015,Kubota2015,Baek2017,Banerjee2017}, which is the field configuration in our experiment.
Remarkably, $\kappa_{xy}$ changes sign at $T_N$ which provides evidence for a significantly different nature of the magnetic quasiparticles in the two phases. In the paramagnetic phase, $\kappa_{xy}$ is positive, exhibits a broad peak at around 30~K, and gradually decays towards higher temperature. The positive $\kappa_{xy}$ in the paramagnetic phase is accompanied by a positive thermal magnetoresistance. Our findings unambiguously prove the unconventional nature of spin excitations in $\alpha$-RuCl$_3$.

For the measurements of the longitudinal and transversal heat conductivity, a single crystal of $\alpha$-RuCl$_3$ was grown using the chemical vapor transport method as described in Ref.~\onlinecite{Hentrich2018}. 
The roughly rectangular shaped crystal was mounted in our home-built heat conductivity setup with the honeycomb planes perpendicular to the magnetic field direction $z$. An in-plane heat current $j_x$ perpendicular to the magnetic field was generated by a chip heater (see inset of Fig.~\ref{kapx} for a sketch of the experimental setup). The resulting longitudinal temperature gradient $\partial _x T$ was measured utilizing a magnetic field calibrated Au/Fe-Chromel thermocouple with the junctions mounted on one of the sample's faces along the $x$-direction (labeled $x$-thermocouple in the figure). A second, transverse thermocouple (labeled $y$-thermocouple) with the junctions attached to opposing crystal edges was used for measuring the transversal temperature gradient $\partial _y T$. In a first step, the longitudinal heat conductivity $\kappa _{xx} (T)$ was measured at zero and finite magnetic fields, applying only a very small heat current.
Since the thermal Hall signal typically is very small, an increased heat current was applied in a second step, in order to obtain a reasonable signal to noise ratio in the thermal Hall effect measurement. To ensure the consistency of the results despite the large thermal perturbation, the heat current applied was varied at several fixed temperature values in order to varify a linear response in $\partial _y T$.

The sample temperature was determined by extrapolating the thermal bath temperature to the $x$-position of the $y$-thermocouple at the sample's center (see inset of Fig.~\ref{kapx}). For this purpose the previously obtained $\kappa _{xx}$ value was used, assuming a constant temperature gradient across the sample. 
For eliminating any longitudinal contribution in the $\partial _y T$-signal due to a possible slight offset of the $y$-thermocouple's contacts in $x$-direction, $\partial _y T (B)$ was measured for both field polarities. The transversal heat conductivity $\kappa_{xy}$ has then been evaluated with the antisymmetrized signal 
$\partial _y ^a T (B) =(\partial _y T (B)-\partial _y T (-B))/2$ as
\begin{equation}
 \kappa_{xy}=\frac{\kappa_{xx}^2{\partial _y ^a T}}{j_x}.
\end{equation}
Both the field dependence $\kappa_{xy} (B)$ at selected fixed temperatures and the temperature dependence $\kappa_{xy} (T)$ at a fixed field of 16~T have been measured. The results for each $\kappa_{xy} (B)$ was then fit assuming a linear field dependence and the resulting 16~T values were used to consolidate the $T$-dependent measurement at fixed field. Each $\kappa_{xy} (B,T)$ was measured multiple times, enabling a rough estimation of the statistical uncertainty of each datapoint.

$\kappa _{xx} (T)$ of $\alpha$-RuCl$_3$ at zero magnetic field is very similar to that of our previous results \cite{Hentrich2018}. The observed $T$-dependence can be rationalized to originate essentially from heat transport by phonons which above $T_N$ experience significant scattering through the Kitaev-Heisenberg paramagnons and thus cause a pertinent suppression of the phononic heat transport.
The minimum at $T_N\approx8$~K and the sharp increase at lower $T$ indicate the abrupt freezing-out of this scattering in the magnetically ordered phase.
Remarkably, in contrast to the previously investigated case where the field was applied parallel to the honeycomb layers, the magnetic field \textit{perpendicular} to the planes has only a weak effect on $\kappa _{xx}(T)$.  As can be inferred from Fig.~\ref{kapx}, at $B _z =16$~T, the thermal conductivity is slightly reduced for $T \lesssim$125~K at most by about 5.5\% at 38~K, with the overall shape being nearly unchanged.

\begin{figure}[t]
\centering
\includegraphics[width=\columnwidth]{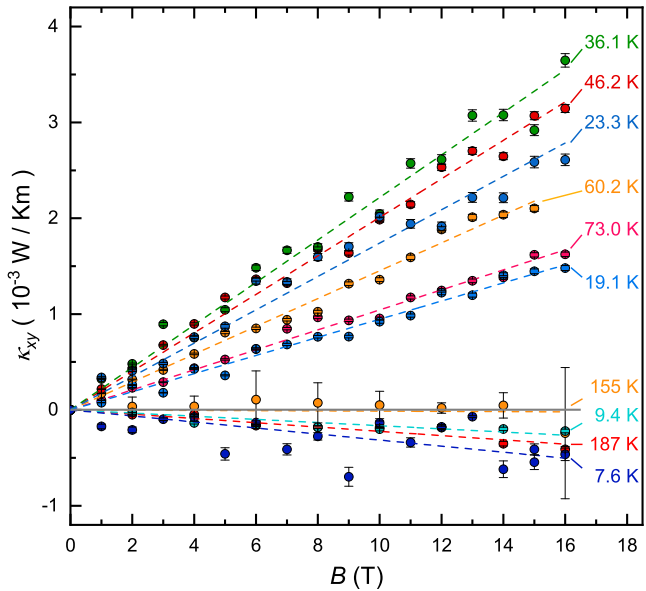}
\caption{\label{kappaBxy}
Field dependence of the transversal heat conductivity $\kappa_{xy}$ of $\alpha$-RuCl$_3$ at selected temperatures. The dashed lines are linear fits to the data which were consecutively used in comparison to the $T$ dependent measurements. Error bars are based on the fit quality with respect to a linear field dependence. }
\end{figure}

Fig. \ref{kappaBxy} shows the results of the field dependent measurements of the transverse thermal conductivity $\kappa_{xy} (B)$ at fixed $T$.
At all measured $T$, the field dependence is linear as emphasized by the linear fits to the data (dashed lines).
In the vicinity of $T_N$, $\kappa_{xy} (B)$ is negative and assumes relatively small negative values of a few $10^{-4}~\rm WK^{-1}m^{-1}$ at about 10~T. However, at somewhat higher $19~{\rm K}\leq T\leq 73$~K, $\kappa_{xy} (B)$ is positive and reaches about one order of magnitude larger values. 

At further elevated $T$, $\kappa_{xy} (B)$ becomes significantly smaller. At $T=$155~K, $\kappa_{xy} (B)$ remains practically zero whereas at even higher $T=187$~K  and $T=230$~K (not shown) small negative and positive values, again in the range of the values for lowest $T$ are obtained, respectively.

\begin{figure}[t]
\centering
\includegraphics[width=\columnwidth]{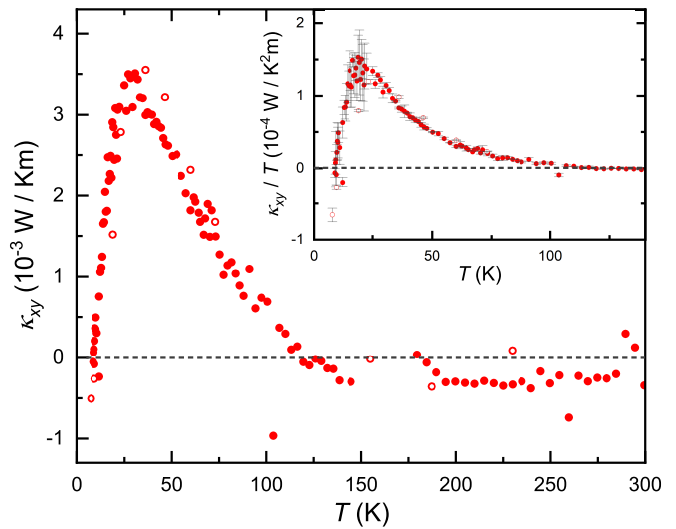}
\caption{\label{kappaTxy}
Temperature dependence of the transversal heat conductivity $\kappa_{xy}$ of $\alpha$-RuCl$_3$ at $|B_z|=16$~T, applied perpendicular to the honeycomb planes and kept fixed during the temperature cycles. The open symbols represent the respective results of the field dependent measurements and show excellent agreement with the $T$ dependent measurements. 
Inset: $\kappa_{xy}/T$ of $\alpha$-RuCl$_3$ as a function of temperature.
}
\end{figure}

In order to shed more light on the $T$-dependence of $\kappa_{xy}$, we present in Fig.~\ref{kappaTxy} $\kappa_{xy}(T)$ measured at a fixed field $|B_z|=16$~T. In agreement with the field dependent data, $\kappa_{xy}(T)$ is small and negative ($\kappa_{xy}\sim -10^{-4}~\rm WK^{-1}m^{-1}$) at $T\lesssim T_N$. Upon increasing $T$ through $T_N$ and further, it increases strongly towards large positive values and exhibits a broad peak centered at about 30~K where $\kappa_{xy}\approx3.5\cdot10^{-3}~\rm WK^{-1}m^{-1}$. $\kappa_{xy}(T)$ then decays gradually at further elevated $T$. At about 125~K, $\kappa_{xy}$ reaches zero and turns to small negative values again where it saturates at about $-3\cdot10^{-4}~\rm WK^{-1}m^{-1}$.

It is important to mention that in the regime $T\gtrsim125$~K, the $B$-dependent data at fixed $T$ suggest $\kappa_{xy}\approx0$ rather than the slightly negative result of the $T$-dependent data at fixed field, see Fig.~\ref{kappaTxy}. Since the former measurement mode is not prone to slight inevitable variations of the thermal contacts of the transverse thermocouple during thermal ramping which hamper the precision of determining $\partial _y ^a T$,  in contrast to the latter, we conjecture that the negative $\kappa_{xy}(T)$ at high $T$ should be understood as a consequence of experimental uncertainty, and we conclude a negligible thermal Hall effect at $T\gtrsim 125$~K, see also inset of Fig.~\ref{kappaTxy} which shows $\kappa_{xy}(T)/T$.
We mention further, that between about 150~K and 180~K it was impossible to achieve reproducible data in the $T$-dependent measurements which we attribute to the proximity to the structural phase transition at about 155~K \cite{Cao2016}.

The observation of a sizeable $\kappa_{xy}$ in the paramagnetic phase of the putative Kitaev-Heisenberg system $\alpha$-RuCl$_3$  is the main observation presented in this paper. 
It provides straightforward evidence for magnetic heat transport in $\alpha$-RuCl$_3$ and underpins the unconventional nature of the magnetic quasiparticles in this compound because these emerge from a state lacking magnetic long range order.
With respect to the Kitaev model, the emergence of a sizeable $\kappa_{xy}$ is a priori not unexpected. Kitaev has shown that in this case chiral edge modes are expected to arise in a magnetic field at very low $T$, resulting in a quantized low-$T$ limit of $\kappa_{xy}/T$ \cite{Kitaev2006,Nasu2017}. This situation, without any doubt, is not realized in the present study, because interactions beyond the Kitaev model lead to the observed long-range magnetic order at $T<T_N$, preempting such physics. I.e., the temperature where our study reveals a large $\kappa_{xy}$ is not directly connected to the low-$T$ limit. 

On the other hand, Nasu et al. have recently theoretically predicted a sizeable
$\kappa_{xy}$ to arise in the pure Kitaev model from itinerant Majorana fermions at
intermediate $k_BT\lesssim K$ (with $K$ the Kitaev interaction) \cite{Nasu2017}. Indeed, if directly
compared to our data, these model calculations yield a peak-like $T$-dependence for $\kappa_{xy}$ similar to our experimental result if one uses
$K/k_B=100$~K, i.e. a Kitaev interaction of the order of experimentally determined values for $\alpha$-RuCl$_3$ \cite{Banerjee2016,Sandilands2015}.  Several comments are in order. First, a
persistent issue of a direct comparison of pure Kitaev models with $\alpha$-RuCl$_3$ is
rooted in the low-$T$ long range magnetic order, which clearly shows that
further interactions beyond the Kitaev model should be expected, such as Heisenberg
and anisotropy terms \cite{Jackeli2009, Chaloupka2010, Chaloupka2013, Kim2015,
Yadav2016, Winter2018}. 
Second, in view of the sign change at $T_N$ it remains to be clarified to what extent the incipient order influences the position of the peak in $\kappa_{xy}(T)$.
This has to be contrasted against Kitaev QSLs, where the downturn in
$\kappa_{xy}(T)$ below the peak is pinned to the flux-proliferation temperature
$T^\star$ \cite{Nasu2017}, which is unrelated to $T_N$.  
Third, as emphasized in
 \cite{Nasu2017}, $\kappa_{xy}\propto B^3$ should be one of the striking features from
the chiral edge mode. For a flux gap of $\Delta_\Phi\sim 0.065\times 100~{\rm K} \sim 6.5$~K,
magnetic fields $B\lesssim 10$~T are within the range of validity of this
claim. However our results from Fig.~\ref{kappaBxy} clearly indicate
$\kappa_{xy}\propto B$, typical for most conventional Hall phenomena, rather than
$B^3$. Finally, Kitaev's treatment of the chiral edge state relies on an effective
model, derived from a flux-free state \cite{Kitaev2006}. The applicability of this
model in the random flux state at the experimentally relevant temperatures
$T\gg\Delta_\Phi$ is unclear.  Thus, it is not settled and deserves further
investigation whether theoretical results for the pure Kitaev model are applicable to
our data.

Apart from the fact of a sizeable positive $\kappa_{xy}$ at intermediate $T$, the sign change near $T_N$ needs to be discussed because it implies that the nature of the heat carrying quasiparticles changes here in a fundamental way. 
While in the long-range ordered ground state well defined spin waves with characteristic low-energy spectral weight are present near the $M$-point, the Kitaev-Heisenberg paramagnon spectrum at $T>T_N$ is rather broad and featureless with strong spectral weight at the $\Gamma$-point \cite{Banerjee2017,Banerjee2018}. Our data therefore suggest that the spin waves lead to a negative $\kappa_{xy}$, while the Kitaev-Heisenberg paramagnons generate the observed strong positive signal. 

The peak-like temperature dependence of the positive $\kappa_{xy}$ of the Kitaev-Heisenberg paramagnons resembles in a remarkable way that of other heat-carrying quasiparticles such as phonons, electrons or other quantum magnetic excitations \cite{Berman,Hess2007}. One might therefore conjecture that the peak in $\kappa_{xy}(T)$ reflects the thermal occupation and the scattering characteristics of the Kitaev-Heisenberg paramagnons at the low-$T$ and high-$T$ edges of the peak, respectively. A theorectical approach which addresses a pertinent analysis is therefore highly desirable. In this context it is also important to elucidate to what extent the finite Hall thermal conductivity implies also a magnetic longitudinal contribution to the heat transport which so far has not been observed experimentally. A qualitative explanation for the lack of such an observation might be connected to the fact that the maximum of $\kappa_{xy}(T)$ is at about the same $T\sim30$~K as that of $\kappa_{xx}(T)$, cf. Figs.~\ref{kapx} and \ref{kappaTxy}. Since it is not far-fetched to assume that a putative magnetic contribution to the longitudinal heat transport, $\kappa_{xx,\mathrm{mag}}$, possesses a $T$-dependence similar to that of $\kappa_{xy}$, this $\kappa_{xx,\mathrm{mag}}$ can be expected to become maximal at about 30~K as well. It would thus remain a subtle effect in $T$-dependent data of $\kappa_{xx}$ and therefore difficult to detect.

In this regard it is interesting, as can be inferred from Fig.~\ref{kapx}, that the application of a large magnetic field of 16~T causes a small but significant suppression of $\kappa_{xx}$ in the same $T$-regime where $\kappa_{xy}$ is positive.  The suppression clearly cannot be attributed to a field-induced increase of the phonon-spin scattering: on the one hand, $B$ perpendicular to the honeycomb layers  has much weaker impact on the magnetic spectrum as compared to $B$ parallel to the planes where at $B\gtrsim 8$~T a substantial spin gap is induced \cite{Baek2017,Hentrich2018}. On the other hand, the field-induced change of $\kappa_{xx}$ for the latter case is limited to relatively low $T\lesssim50$~K, whereas for $B_z=16$~T perpendicular to the layers the suppression is detectable up to about 125~K. Thus, one can directly conclude that the observed suppression of $\kappa_{xx}$ in the present case has to be attributed to a finite magnetic contribution $\kappa_{xx,\mathrm{mag}}$ which apparently experiences a positive thermal magnetoresistance effect.

Finally, one should note, that our analysis does not include the consideration of a thinkable phononic thermal Hall effect which has recently been proposed to potentially arise in particular magnetic systems as a secondary effect from spin-phonon coupling \cite{Strohm2005,Sugii2017}. It remains subject to future investigations to what extent phonons contribute to the sizeable $\kappa_{xy}$ in $\alpha$-RuCl$_3$ in addition to the magnetic excitations.

In conclusion, we have observed a sizeable magnetic thermal Hall effect in the Kitaev-Heisenberg material $\alpha$-RuCl$_3$ in magnetic fields up to 16~T perpendicular to the honeycomb layers of the compound. The transverse heat conductivity is linear in the magnetic field, i.e., $\kappa_{xy}(B)\propto B$ at all temperatures. $\kappa_{xy}$ is relatively small and negative in the magnetically ordered phase at $T\lesssim T_N\approx8$~K, whereas it is positive in the paramagnetic phase up to  $T\approx125$~K with a large peak at about 30~K. The occurrence of the positive  $\kappa_{xy}$ is accompanied by a significant positive thermal magnetoresistance in the longitudinal heat conductivity $\kappa_{xx}$. Altogether, our results provide clear-cut evidence for heat transport of the putative Kitaev-Heisenberg paramagnons in $\alpha$-RuCl$_3$, which underpins the exotic nature of these quasiparticles.

Final note: upon finalizing our work we became aware of another study of $\kappa_{xy}$ in $\alpha$-RuCl$_3$ by Kasahara et al. \cite{Kasahara2018} with similar experimental results. We mention, however, that the reported non-linear $B$-dependence of $\kappa_{xy}$ is not supported by our data.

\begin{acknowledgments}
This work has been supported by the Deutsche Forschungsgemeinschaft through SFB 1143 (projects A02, B03, and C07) and through the projects HE3439/12 and HE3439/13. W.B. acknowledges partial support by QUANOMET, CiNNds, and PSM.

\end{acknowledgments}



%

\end{document}